\documentclass{ptephy_v1}
\usepackage{amsmath,amssymb}
\usepackage{color}
\usepackage{mathrsfs}
\usepackage{subcaption}

\preprintnumber{CHIBA-EP-249} 

\begin{document}

\title{Center group dominance in quark confinement}

\author{Ryu Ikeda}
\affil{Department of Physics, Graduate School of Science and Engineering, Chiba University, Chiba 263-8522, Japan\email{cdna0955@chiba-u.jp}}

\author[1,2]{Kei-Ichi Kondo}
\affil{Department of Physics, Graduate School of Science, Chiba University, Chiba 263-8522, Japan\email{kondok@faculty.chiba-u.jp}}


\begin{abstract}%

We show that the color $N$ dependent area law falloffs of the double-winding Wilson loop averages for the $SU(N)$ lattice gauge theory obtained in the preceding works are reproduced from the corresponding lattice Abelian gauge theory with the center gauge group $Z_N$.  
This result indicates the center group dominance in quark confinement. 

\end{abstract}

\subjectindex{B64 Lattice QCD}

\maketitle


\noindent
\textit{1. Introduction.} \ 
The area law falloff of the Wilson loop average is the most well-known criterion for quark confinement, which implies the existence of a linear potential between a static quark-antiquark pair in the gauge-independent way \cite{Wilson74}. For recent developments on quark confinement, see e.g., \cite{Greensite03,KKSS15} for reviews. 
Quite recently, the double-winding Wilson loop has been introduced in the lattice gauge theory by Greensite and H\"{o}llwieser \cite{GH15} to examine the possible mechanisms for quark confinement. 
In the $SU(2)$ lattice gauge theory, in particular, a double-winding Wilson loop was used as a probe to compare two promising pictures for quark confinement and thereby single out the true confinement mechanism according to the area law falloff of its average:
the sum-of-areas law expected in the Abelian magnetic monopole picture and the difference-of-areas law expected in the center vortex picture.

The \textit{double-winding Wilson loop operator} $W( C_1 \cup C_2 )$ is a trace of the path-ordered product of gauge link variables $U_\ell$ along a closed loop $C$ composed of two loops $C_1$ and $C_2$:
\begin{align}
W( C_1 \cup C_2 ) \equiv {\rm tr} \left[ \prod_{\ell \in C_1 \cup C_2}U_\ell \right]. 
\label{dw-Wilson-loop}
\end{align}

\begin{figure}[!h]
\centering\hspace{-10mm}
\begin{subfigure}{70mm}
  \centering\includegraphics[width=70mm]{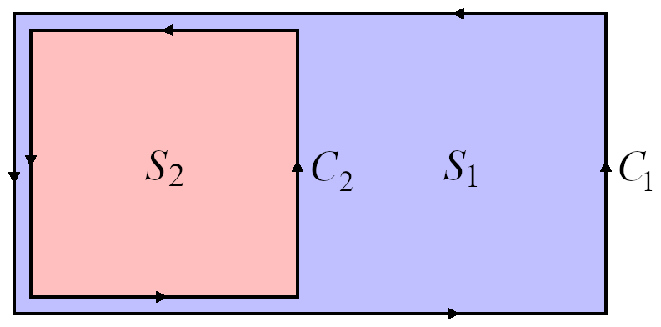}
  \hspace{-10mm} \caption{  }
  \label{fig1}
\end{subfigure}
\begin{subfigure}{75mm}
  \centering\includegraphics[width=75mm]{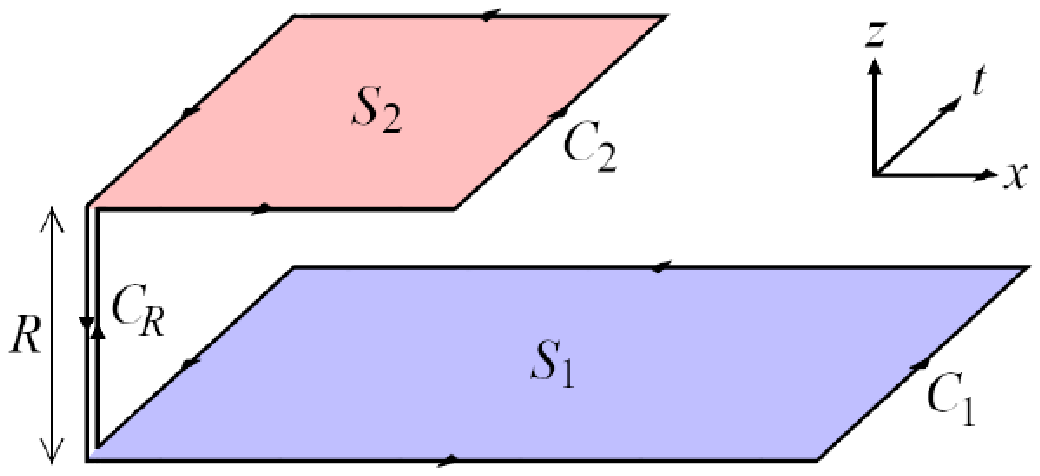}
  \hspace{-10mm} \caption{  }
  \label{fig2}
\end{subfigure}
\hspace{-10mm} \caption{(a) a ``coplanar'' double-winding Wilson loop, (b) a ``shifted'' double-winding Wilson loop composed of the two loops $C_1$ and $C_2$ which lie in planes parallel to the $x-t$ plane, 
but are displaced from one another in the $z$-direction by distance $R$.}
\label{fig:dw-Wilson-loop}
\end{figure}

The double-winding Wilson loop is called \textit{coplanar} if the two loops $C_1$ and $C_2$ lie in the same plane, while it is called \textit{shifted} if the two loops $C_1$ and $C_2$ lie in planes parallel to the $x-t$ plane, but are displaced from one another in the transverse direction, e.g., $z$ by distance $R$, and are connected by lines running parallel to the $z$-axis to keep the gauge invariance. 
The coplanar case is regarded as a non-shifted $R=0$ limit of the shifted one. 
Note that the double-winding Wilson loop operators are defined in a gauge invariant manner, irrespective of shifted $R \not= 0$ or coplanar $R=0$.
See Fig.\ref{fig:dw-Wilson-loop}.
From the physical viewpoint, a double-winding Wilson loop operator represents the correlation between the two interacting quarkonia, namely, two interacting mesons which consist of a heavy quark and its antiquark. It can be used as  a probe to study the interactions between two flux tubes connecting quark-antiquark pairs. 


The preceding works \cite{GH15,MK17,KSK20} investigated the area   ($S_1$ and $S_2$) dependence of the expectation value $\left< W(C_1 \cup C_2) \right>$ of a double-winding Wilson loop operator $W(C_1 \cup C_2)$
where 
$S_1$ and $S_2$ are respectively the minimal areas bounded by loops $C_1$ and $C_2$.
In the lattice $SU(2)$ Yang-Mills theory, it has been first shown in \cite{GH15}  that the coplanar double-winding Wilson loop average obeys the  ``difference-of-areas law''  by using the strong coupling expansion and the numerical simulations on the lattice.

In the continuum $SU(N)$ Yang-Mills theory, subsequently, a double-winding, a triple-winding, and general multiple-winding Wilson loops are investigated  in \cite{MK17} to show  
that a coplanar double-winding $SU(3)$ Wilson loop average follows a novel area law which is neither difference-of-areas law nor sum-of-areas law, and that ``sum-of-areas law'' is allowed for $SU(N)$ ($N \geqslant 4$), provided that the string tension obeys the Casimir scaling for quarks in the higher representations.

In the lattice $SU(N)$ Yang-Mills theory, moreover, it has been shown in \cite{KSK20} by using the strong coupling expansion and the numerical simulations on the lattice that 
the coplanar double-winding Wilson loop average has the $N$ dependent area law falloff:  
``max-of-areas law'' for $N=3$ and ``sum-of-areas law'' for $N \geqslant 4$. 
Moreover, a shifted double-winding Wilson loop average as a function of the distance in a transverse direction has the long distance behavior which does not depend on the number of color $N$, while the short distance behavior depends on $N$.

In order to give better understanding on these results, in this paper, we investigate the area law falloff of the double-winding Wilson loops using the lattice gauge theory with the gauge group $Z_N$, which is the center group of the original color gauge group $SU(N)$. 
We show that the $N$ dependent area law falloffs of the double-winding $SU(N)$ Wilson loop averages can be reproduced by using the Abelian gauge theory with the corresponding center group $Z_N$.
These results show that the center group as a subgroup of the original non-Abelian gauge group has the essential contribution for quark confinement. 
For obtaining the explicit expression of the double-winding Wilson loop average, we make use of the character expansion \cite{Creutz83,MM94} to rewrite the weight coming from the action and perform the $Z_N$ group integration.

Indeed, such dominance of the center group in quark confinement was shown long ago by Fr\"{o}hlich in \cite{Frohlich79} which states that confinement in $Z_N$ lattice gauge theories implies confinement in $SU(N)$ lattice Higgs theories.
According to Fr\"{o}hlich's original argument, we can extend the center group dominance to an arbitrary closed loop composed of any number of loops. Therefore, Fr\"{o}hlich's original result  holds not only for the case of the single-winding Wilson loop average but also for the case of the double-winding and more general multiple-winding Wilson loop averages, beyond the ordinary single-winding Wilson loop average. 
Our calculations of the double-winding Wilson loop average to be performed in the approximation up to the leading order are consistent with this rigorous result, which reinforces the validity of our result of approximate calculations.
In the $SU(N)$ theory, it is an involved task to evaluate the double-winding Wilson loop average, due to the non-Abelian nature of the gauge group $SU(N)$. Applying the Fr\"{o}hlich's argument, we can avoid this complexity to obtain the qualitative properties of the double-winding Wilson loop operator in the $SU(N)$ theory owing to the simplicity of the $Z_N$ theory.\\

\begin{figure}[!h]
\centering
\begin{subfigure}{75mm}
  \centering\includegraphics[width=75mm]{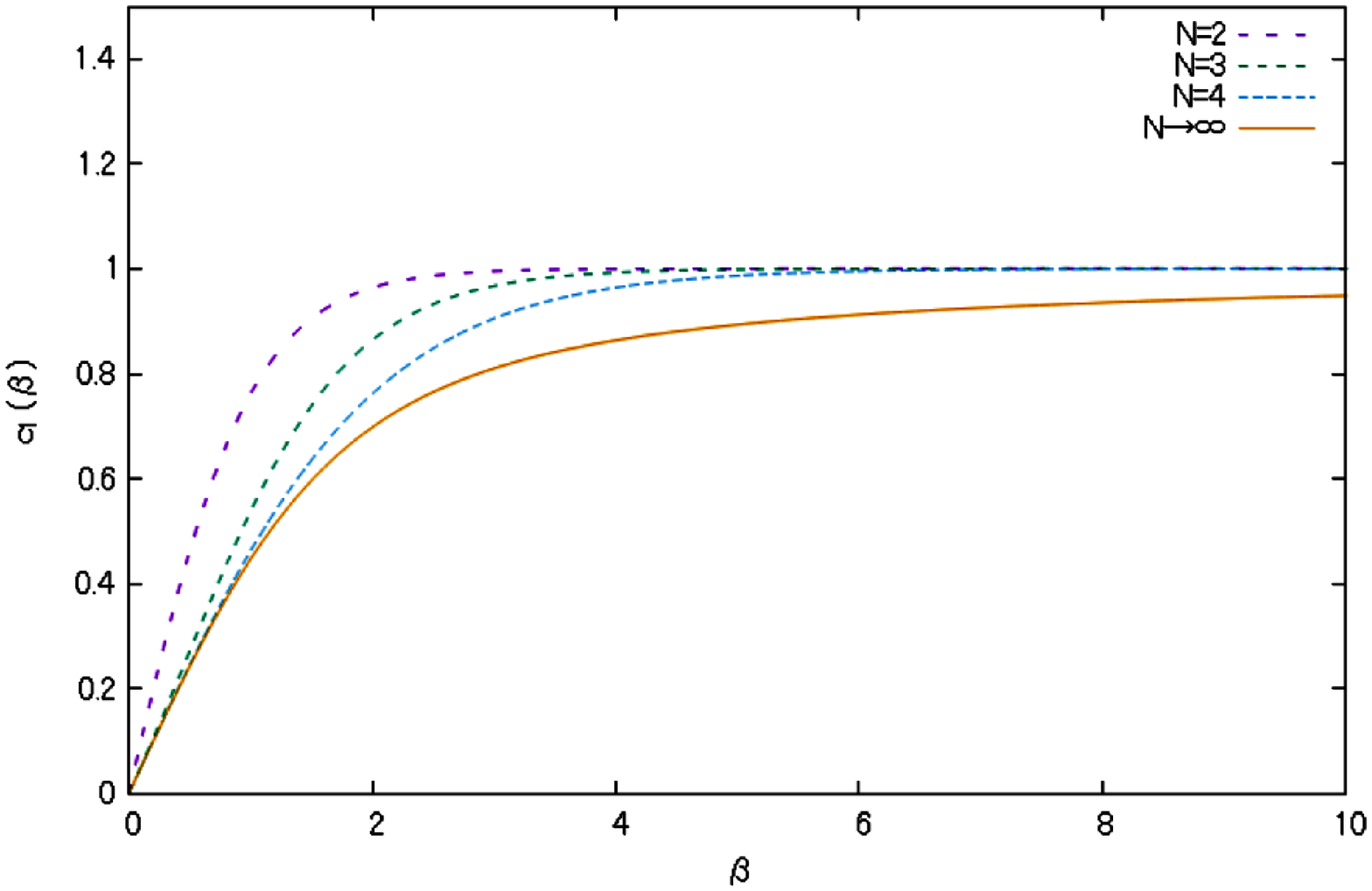}
  \caption{$c_1 (\beta) \ (N=2,3,4,\infty)$}
  \label{fig-f}
\end{subfigure}
\begin{subfigure}{75mm}
  \centering\includegraphics[width=75mm]{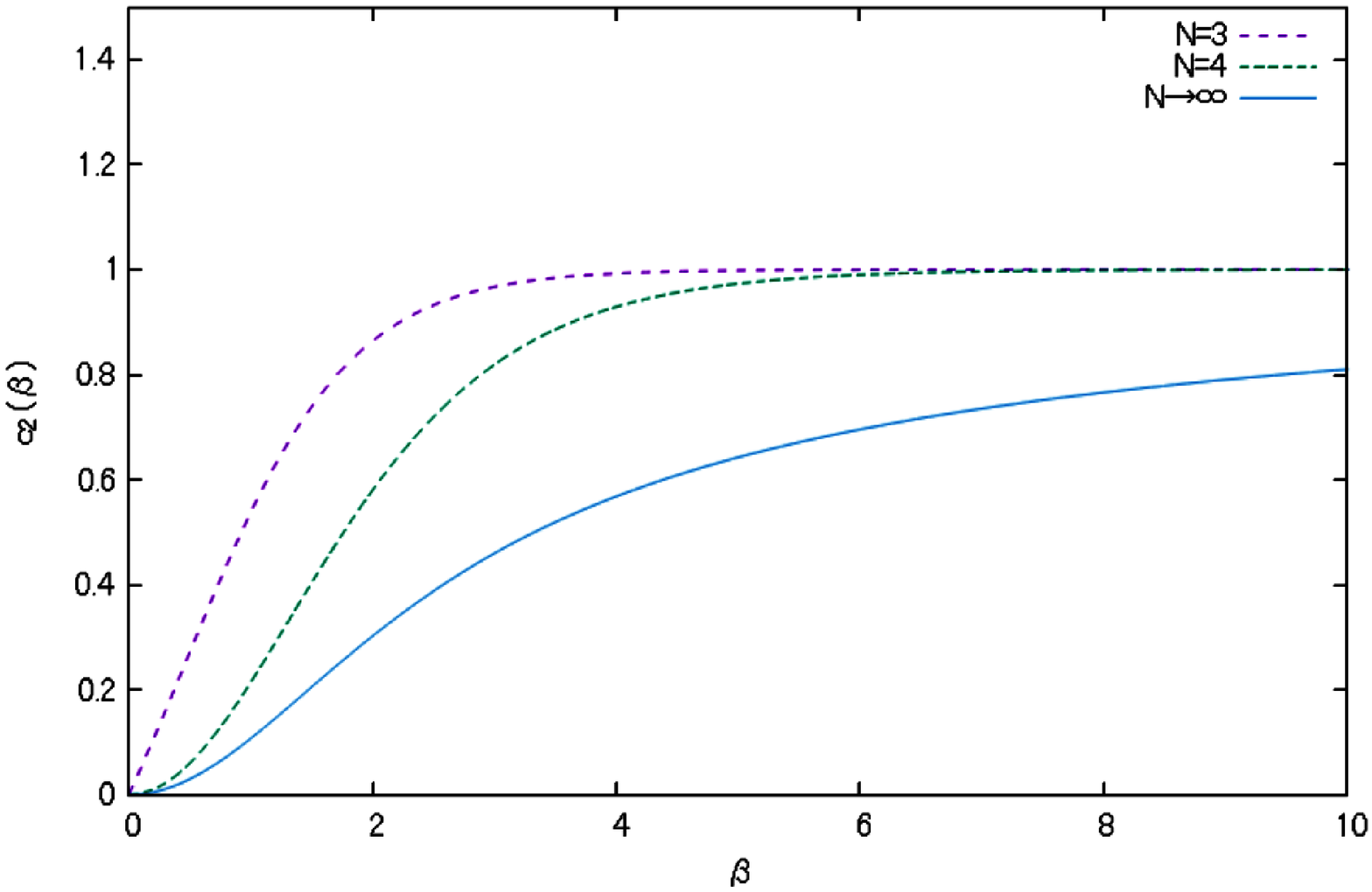}
  \caption{$c_2 (\beta) \ (N=3,4,\infty)$}
  \label{fig-g}
\end{subfigure}
\caption{The character expansion coefficient as a function of $\beta$, (a) $c_1 (\beta)$, \ (b) $c_2 (\beta)$}
\label{fig-fg}
\end{figure}

\noindent
\textit{2. The lattice $Z_N$ pure gauge model.} \ 
First, we introduce the lattice $Z_N$ pure gauge model with the coupling constant defined by $\beta := 1/ g^2$. The action of this model on a $D$-dimensional hypercubic lattice $\Lambda$ with unit lattice spacing is given by
\begin{equation}
  S_G [U] = \beta \sum_{p \in \Lambda} \mathrm{Re} \ U_p , \quad U_p := \prod_{\ell \in \partial p} {U}_{\ell} ,
\end{equation}
where $\ell$ labels a link, $p$ labels an elementary plaquette and $U_p$ is a plaquette variable defined by the path-ordered product of the link variables ${U}_{\ell}$ along the loop $\partial p$, where ${U}_{\ell}$ is a $Z_N$ link variable on link $\ell$. To examine this $Z_N$ model analytically, we introduce the character expansion. We apply the \textit{character expansion} to the weight ${e}^{ S_G [U] }$ to obtain
\begin{equation}
  {e}^{ S_G [U] }
    = \prod_{p \in \Lambda} {e}^{\beta \mathrm{Re} \ U_p} = \prod_{p \in \Lambda} \sum_{n=0}^{N-1} b_n (\beta) U_p^n ,
\end{equation}
where $b_n (\beta)$ is defined by
\begin{equation}
  b_n (\beta) := \frac{1}{N} \sum_{\zeta \in Z_N} {\zeta}^{-n} {e}^{\beta \mathrm{Re} \ \zeta}.
\end{equation}
Then the expectation value of an operator $\mathscr{F}$ is given by
\begin{align}
  {\langle \mathscr{F} \rangle}_{\Lambda}
    &= {Z}_{\Lambda}^{-1} \int \prod_{\ell \in \Lambda} d {U}_{\ell} \ {e}^{ S_G [U] } \mathscr{F}
    = {Z}_{\Lambda}^{-1} \int \prod_{\ell \in \Lambda} d {U}_{\ell} \ \prod_{p \in \Lambda} \sum_{n=0}^{N-1} b_n (\beta) U_p^n \mathscr{F}, \\
  {Z}_{\Lambda}
    &:= \int \prod_{\ell \in \Lambda} d {U}_{\ell} \ {e}^{ S_G [U] }.
\end{align}
The $Z_N$ link variable ${U}_{\ell} = \exp \left( i 2 \pi {k}_{\ell} /N \right) \ ({k}_{\ell} = 0,1, \cdots ,N-1)$ in the continuous group limit $N \to \infty$ reduces to $U(1)$ link variable ${U}_{\ell} = \exp \left( i {\theta}_{\ell} \right) \ (-\pi < {\theta}_{\ell} \leqslant \pi)$.\\
For the action of the lattice $U(1)$ pure gauge model given by
\begin{equation}
  S_G [U] = \beta \sum_{p \in \Lambda} \mathrm{Re} \ U_p = \beta \sum_{p \in \Lambda} \cos {\theta}_{p} , \quad {\theta}_{p} := \sum_{\ell \in \partial p} {\theta}_{\ell},
\end{equation}
and the character expansion of  the weight ${e}^{S_G [U]}$ is given by
\begin{equation}
  {e}^{ S_G [U] }
    = \prod_{p \in \Lambda} {e}^{\beta \cos {\theta}_{p}} = \prod_{p \in \Lambda} \sum_{n=0}^{\infty} b_n (\beta) {e}^{in {\theta}_{p} }.
\end{equation}
Notice that $b_n (\beta)$ for the continuous group $U(1)$ agrees with the integral representation of the first-kind modified Bessel function $I_n (\beta)$:
\begin{align}
  b_n (\beta)
    := \frac{1}{2\pi} \int_{-\pi}^{\pi} d\theta \ {e}^{-in\theta} {e}^{\beta \cos \theta}
    = \frac{1}{\pi} \int_{0}^{\pi} d\theta \ \cos (n\theta) {e}^{\beta \cos \theta}
    =: I_n (\beta).
\end{align}
We define $c_n (\beta)$ by $c_n (\beta) = b_n (\beta) / b_0 (\beta)$. For $N=2,3,4$ and $\infty$, $c_1 (\beta)$ and $c_2 (\beta)$ are written in the form
\begin{align}
  c_1 (\beta) &= \frac{ {e}^{\beta} - {e}^{-\beta} }{ {e}^{\beta} + {e}^{-\beta} } \quad (N=2) \ , \qquad 
  c_1 (\beta) = \frac{ {e}^{\beta} - {e}^{-\beta/2} }{ {e}^{\beta} + 2{e}^{-\beta/2} } = c_2 (\beta) \quad (N=3) \ , \notag \\
  c_1 (\beta) &= \frac{ {e}^{\beta} - {e}^{-\beta} }{ {e}^{\beta} +2+ {e}^{-\beta} } , \quad
  c_2 (\beta) = \frac{ {e}^{\beta} -2+ {e}^{-\beta} }{ {e}^{\beta} +2+ {e}^{-\beta} } \quad (N=4) \ , \notag \\
  c_1 (\beta) &= \frac{ I_1 (\beta) }{ I_0 (\beta) } , \quad
  c_2 (\beta) = \frac{ I_2 (\beta) }{ I_0 (\beta) } \quad (N=\infty).
\end{align}
Note that ${b}_{N-n} (\beta) = b_n (\beta)$ and $0 \leqslant c_n (\beta) < 1$ for $0 \leqslant \beta < \infty$. For $N=2,3,4$ and $\infty$ , the behavior of $c_1 (\beta)$ and $c_2 (\beta)$ as functions of $\beta$ are indicated in Fig.\ref{fig-fg}. We find that $c_1 (\beta) \sim \mathcal{O} (\beta)$ and $c_2 (\beta) \sim \mathcal{O} ({\beta}^{2})$ for $\beta \ll 1$, while $c_1 (\beta) \sim c_2 (\beta)$ for $\beta \gg 1$, and $c_1 (\beta), c_2 (\beta) \to 1$ as $\beta \to \infty$, irrespective of $N$.

\noindent
\textit{3. The evaluation of Wilson loop averages in the lattice $Z_N$ pure gauge model.} \ 
Next, we evaluate the expectation values of an ordinary single-winding Wilson loop, a coplanar double-winding Wilson loop, and a shifted double-winding Wilson loop in the lattice $Z_N$ pure gauge model. 

\noindent
\textit{3.1. The Wilson loop average in the $Z_N$ pure gauge model.} \ 
For a link variable ${U}_{\ell} \in Z_N$, ${U}_{\ell}$ is written as ${U}_{\ell} = \exp \left( i 2 \pi {k}_{\ell} /N \right) \ ({k}_{\ell} = 0,1, \cdots ,N-1)$, which is a complex number of unit modulus, ${U}_{\ell} \in \mathbb{C} , \ |{U}_{\ell}| = 1$.
Therefore, ${({U}_{\ell})}^{N} =1$ and ${({U}_{\ell}^{*})}^{n} = {({U}_{\ell})}^{N-n}$. The integration formulae for the $Z_N$ group is very simple:
\begin{align}
\label{int}
  \int dU \ {(U)}^{m} {(U^*)}^{n} = {\delta}_{mn} \qquad (m,n \in \mathbb{Z}, \mod N).
\end{align}

\begin{figure}[!h]
\centering
\begin{subfigure}{75mm}
  \centering\includegraphics[width=75mm]{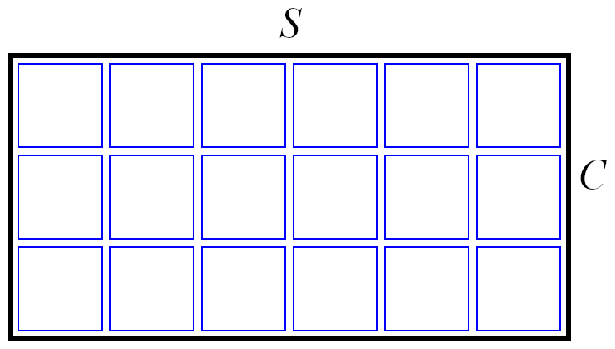}
  \hspace{-10mm} \vspace{3mm} \caption{$N=2$}
  \label{fig-0}
\end{subfigure}
\begin{subfigure}{75mm}
  \centering\includegraphics[width=75mm]{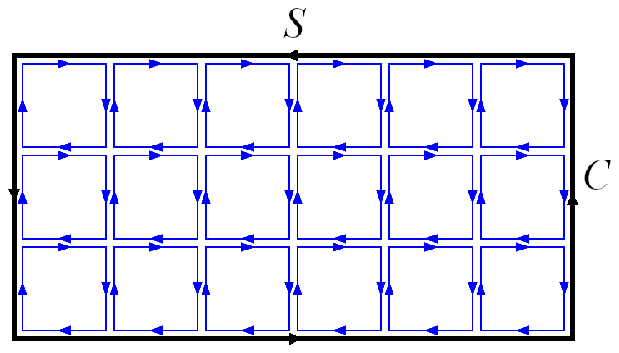}
  \hspace{-10mm} \vspace{3mm} \caption{$N \geqslant 3$}
  \label{fig-a}
\end{subfigure}
\caption{A single-winding Wilson loop, (a) $N=2$ , \ (b) $N \geqslant 3$}
\label{fig-0a}
\end{figure}

In a Wilson loop operator $W(C) := \prod_{\ell \in C} {U}_{\ell}$, there is a single link variable ${U}_{\ell}$ on a link $\ell \in C$. For obtaining non-vanishing contributions after integration, there must be another link variable ${U}_{\ell}^{*} = {U}_{\ell}^{N-1}$ for ${U}_{\ell} \ (\ell \in C)$ which is supplied from the character expansion of the weight ${e}^{S_G [U]}$.
It should be remarked that for an Abelian group has a special property:
\begin{equation}
\label{special}
  \prod_{\ell \in C} {U}_{\ell} = \prod_{\substack{p \in S \\ \partial S = C}} {U}_{p} ,
\end{equation}
where $S$ is an arbitrary surface composed of a connected set of plaquettes and bounded by a loop $C$. From this special property of the Abelian group, a Wilson loop operator is converted to the product of $U_p$.
Therefore, the leading contribution is given by the diagram in Fig.\ref{fig-0a} where the minimal area $S$ bounded by the loop $C$ is tiled by a planar set of plaquettes. For $N=2$, given in the left panel of Fig.\ref{fig-0a}, especially, the link variable ${U}_{\ell}$ is real (${U}_{\ell} = \pm 1$) and hence no orientation.

We take a lattice $\Lambda$ sufficiently large so as to include these gauge-invariant operators 
 and calculate the leading contribution.
Therefore the leading contribution for the Wilson loop average $\langle W(C) \rangle$ is calculated as
\begin{align}
\label{wc}
  \langle W(C) \rangle
    &\simeq {Z}_{S}^{-1} \int \prod_{\ell \in S} d {U}_{\ell} \prod_{p \in S} \left[ {b}_{N-1} {U}_{p}^{N-1} + {b}_{N-2} {U}_{p}^{N-2} + \cdots + b_1 U_p + b_0 \right] \prod_{\ell' \in C} {U}_{\ell'} \notag \\
    &= {Z}_{S}^{-1} \int \prod_{\ell \in S} d {U}_{\ell} \prod_{p \in S} \left[ {b}_{N-1} {U}_{p}^{N-1} + {b}_{N-2} {U}_{p}^{N-2} + \cdots + b_1 U_p + b_0 \right] \prod_{p' \in S} {U}_{p'} \notag \\
    &= {Z}_{S}^{-1} \int \prod_{\ell \in S} d {U}_{\ell} \prod_{p \in S} \left[ {b}_{N-1} + {b}_{N-2} {U}_{p}^{N-1} + \cdots + b_1 U_p^2 + b_0 U_p \right],
\end{align}
where the partition function $Z_S$ for $\Lambda = S$ reads
\begin{equation}
  Z_S \simeq \int \prod_{\ell \in S} d {U}_{\ell} \prod_{p \in S} \left[ b_0 + {b}_{N-1} {U}_{p}^{N-1} + \cdots + b_2 U_p^2 + b_1 U_p \right].
\end{equation}
Here we have used $\prod_{\ell' \in C} {U}_{\ell'} = \prod_{p' \in S} {U}_{p'}$ following from (\ref{special}) in the second equality of (\ref{wc}) and the integration formula obtained from (\ref{int}):
\begin{equation}
\label{intp}
  \int \prod_{\ell \in p \cup p'} d {U}_{\ell} \ {(U_p)}^{m} {(U_p')}^{n}
  =\begin{cases}
        \int \prod_{\ell \in p} d {U}_{\ell} \ {(U_p)}^{m+n} &(p = p') \\
        0 &(p \neq p')
    \end{cases}
  \qquad (n \not\equiv 0 \mod N).
\end{equation}
Furthermore, we use the integration formula following from (\ref{int}): for the integration on a link $\ell \in S \setminus C$,
\begin{equation}
  \int d {U}_{\ell} \ {({U}_{\ell})}^{{m}_{\ell}} {({U}_{\ell}^{*})}^{{n}_{\ell}} = {\delta}_{{m}_{\ell} {n}_{\ell}} \qquad ({m}_{\ell} , {n}_{\ell} \in \mathbb{Z} \ , \mod N),
\end{equation}
and for the integration on a link $\ell \in C$,
\begin{equation}
  \int d {U}_{\ell} \ {({U}_{\ell})}^{{m}_{\ell}} = {\delta}_{{m}_{\ell} 0} \ , \quad \int d {U}_{\ell} \ {({U}_{\ell}^{*})}^{{n}_{\ell}} = {\delta}_{0 {n}_{\ell}} \qquad ({m}_{\ell} , {n}_{\ell} \in \mathbb{Z} \ , \mod N).
\end{equation}
Therefore, the non-zero contribution in the integration is given by terms with ${m}_{\ell} = {n}_{\ell} = 0$, namely terms without $U_p$:
\begin{equation}
  \langle W(C) \rangle
    \simeq \frac{ {{b}_{N-1} (\beta)}^{|S|} \int \prod_{\ell \in S} d {U}_{\ell} }{ {b_0 (\beta)}^{|S|} \int \prod_{\ell \in S} d {U}_{\ell} }
    = {{c}_{N-1} (\beta)}^{|S|} = {c_1 (\beta)}^{|S|} \qquad (N \geqslant 2),
\end{equation}
where $|S|$ is the total number of plaquettes on $S$. This leading result does not have the dependence on the dimensionality $D$ and gives the exact result for $D=2$.\\

\begin{figure}[!h]
\centering
\begin{subfigure}{75mm}
  \centering\includegraphics[width=75mm]{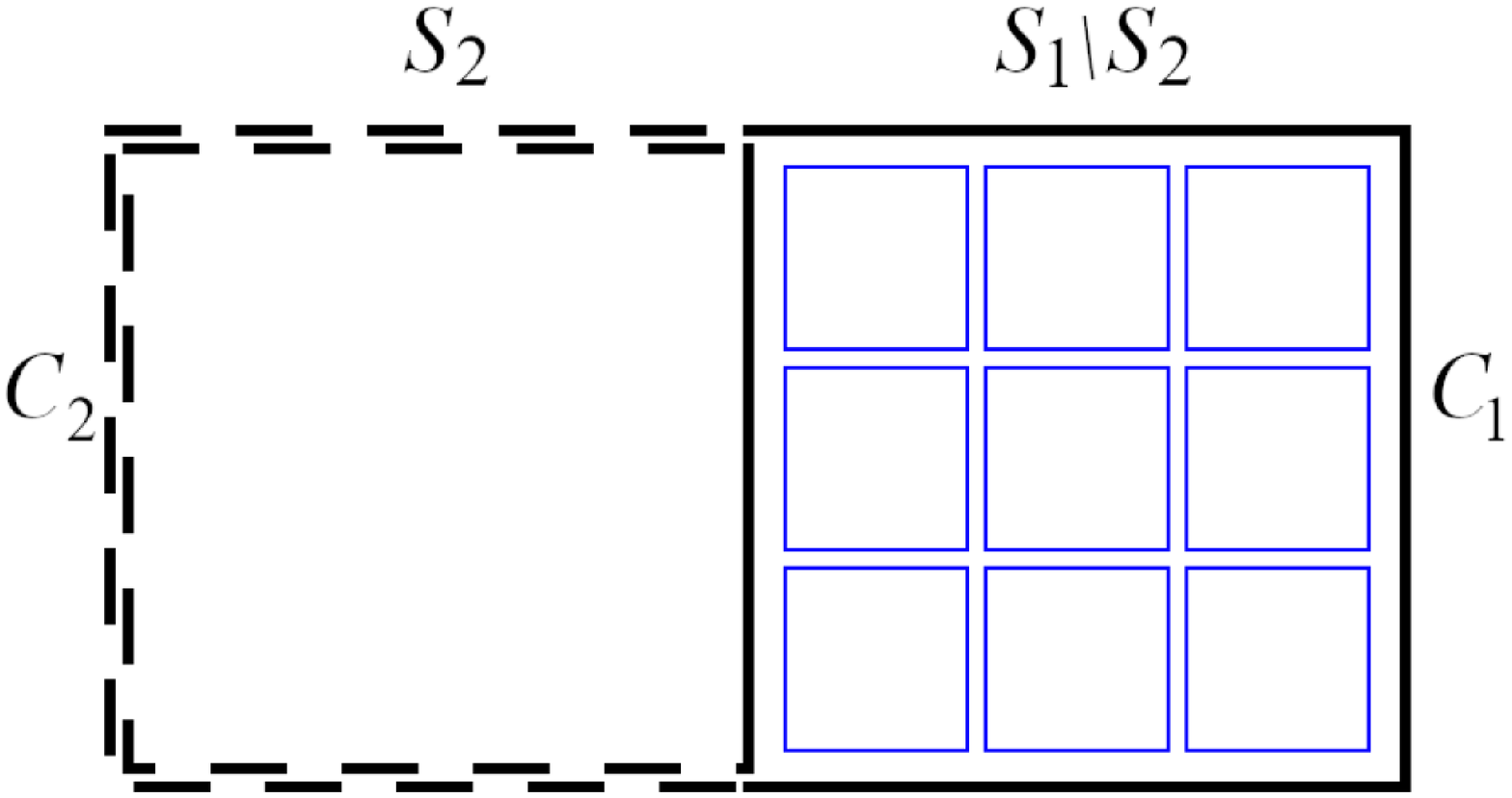}
  \hspace{-5mm} \caption{$N=2$}
  \label{fig-b}
\end{subfigure}
\begin{subfigure}{75mm}
  \centering\includegraphics[width=75mm]{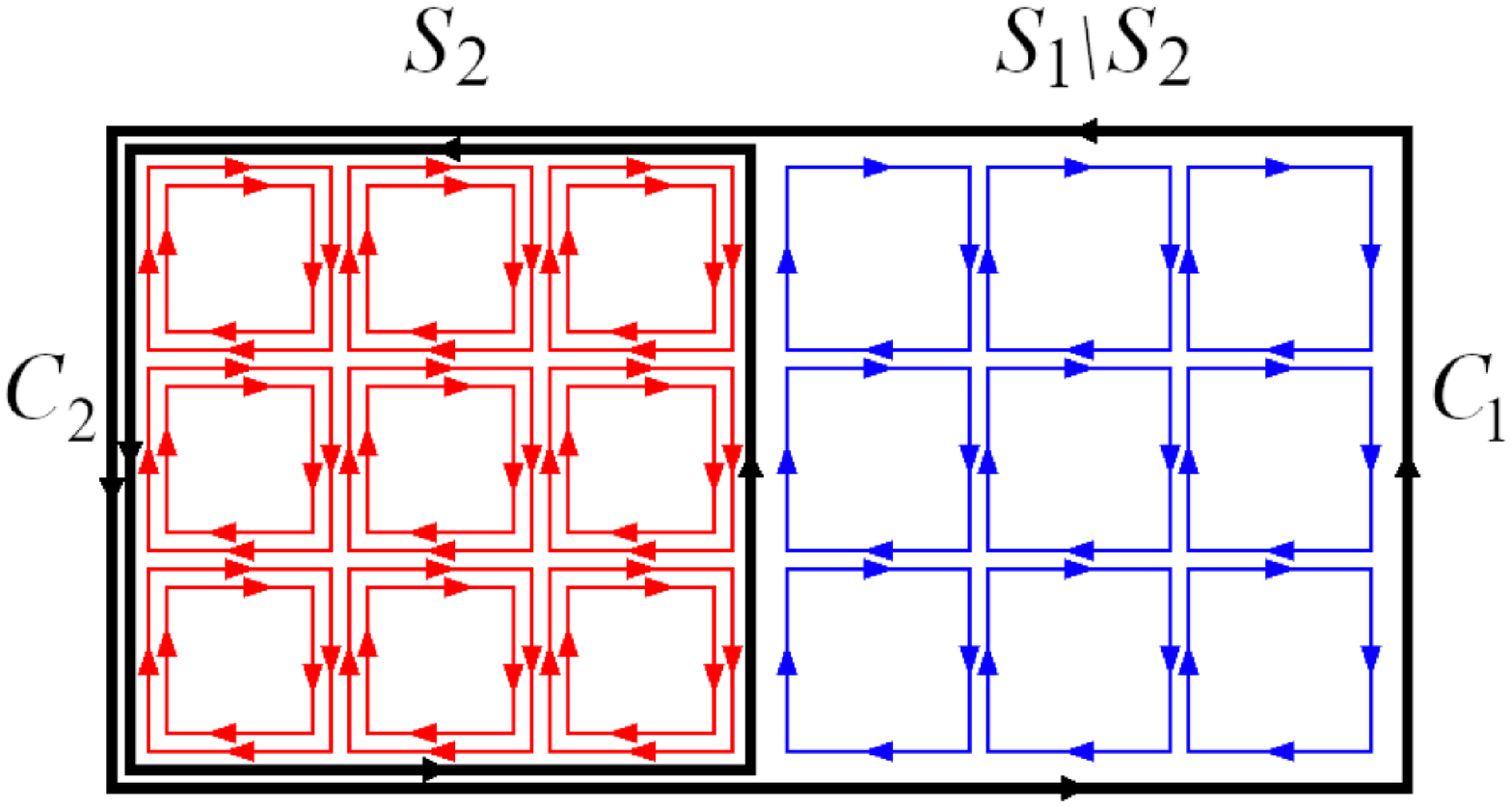}
  \hspace{-5mm} \caption{$N \geqslant 3$}
  \label{fig-c}
\end{subfigure}
\caption{A coplanar double-winding Wilson loop, (a) $N=2$ , \ (b) $N \geqslant 3$}
\label{fig-bc}
\end{figure}

\noindent
\textit{3.2. A coplanar double winding Wilson loop average in the $Z_N$ pure gauge theory.} \ 
From now on, we proceed to study a coplanar double-winding Wilson loop $W(C_1 \cup C_2) := \prod_{\ell \in C_1 \cup C_2} {U}_{\ell}$, where there is a single link variable ${U}_{\ell}$ for $\ell \in C_1 \setminus (C_1 \cap C_2)$ and a double link variable ${U}_{\ell}^{2}$ for $\ell \in C_1 \cap C_2$. See Fig.\ref{fig-bc}. For obtaining non-vanishing contribution after integration, there must be more link variables ${U}_{\ell}^{*} = {U}_{\ell}^{N-1}$ for $\ell \in C_1 \setminus (C_1 \cap C_2)$, and ${({U}_{\ell}^{*})}^{2} = {U}_{\ell}^{N-2}$ for $\ell \in C_1 \cap C_2$, which come from the character expansion of the weight ${e}^{S_G [U]}$.

For $N=2$, ${U}_{\ell}$ is a real number taking the value ${U}_{\ell} = \pm 1$. Therefore, the double link variables ${({U}_{\ell})}^{2} \ (\ell \in C_1 \cap C_2)$ have no contribution due to ${({U}_{\ell})}^{2} =1$. Consequently, a double-winding Wilson loop operator $W(C_1 \cup C_2)$ reduces to a single-winding Wilson loop operator on the loop $C := (C_1 \cup C_2) \setminus (C_1 \cap C_2)$. The leading contribution is given by the diagram in the left panel of Fig.\ref{fig-b} where the area $S_1 \setminus S_2$ is tiled by a planar set of plaquettes. Thus, the leading contribution for the $\langle W(C_1 \cup C_2) \rangle$ is calculated as the difference-of-area law:
\begin{equation}
  \langle W(C_1 \cup C_2) \rangle = \langle W(C) \rangle \simeq {c_1 (\beta)}^{|S_1|-|S_2|}, \quad C := (C_1 \cup C_2) \setminus (C_1 \cap C_2).
\end{equation}

For $N \geqslant 3$, the leading contribution is given by the diagram in the right panel of Fig.\ref{fig-bc} where the area $S_1 \setminus S_2$ is tiled once and $S_2$ is tiled twice by a set of plaquettes which are supplied from the character expansion of the weight ${e}^{S_G [U]}$. Therefore, the leading contribution for $\langle W(C_1 \cup C_2) \rangle$ is calculated as
\begin{align}
  \langle W&(C_1 \cup C_2) \rangle
    \simeq {Z}_{S_1}^{-1} \int \prod_{\ell \in S_1} d {U}_{\ell} \prod_{p \in S_1} \left[ {b}_{N-1} {U}_{p}^{N-1} + {b}_{N-2} {U}_{p}^{N-2} + \cdots + b_1 U_p + b_0 \right] \prod_{\ell' \in C_1 \cup C_2} {U}_{\ell'} \notag \\
    &= {Z}_{S_1}^{-1} \int \prod_{\ell \in S_1} d {U}_{\ell} \prod_{p \in S_1} \left[ {b}_{N-1} {U}_{p}^{N-1} + {b}_{N-2} {U}_{p}^{N-2} + \cdots + b_1 U_p + b_0 \right] \prod_{p' \in S_2} {U}_{p'}^{2} \prod_{p' \in S_1 \setminus S_2} {U}_{p'} \notag \\
    &= {Z}_{S_1}^{-1} \int \prod_{\ell \in S_1} d {U}_{\ell} \prod_{p \in S_2} \left[ {b}_{N-2} + {b}_{N-3} {U}_{p}^{N-1} + \cdots + b_0 U_p^2 + {b}_{N-1} U_p \right] \notag \\
    &\qquad \qquad \qquad \qquad \qquad \times \prod_{p \in S_1 \setminus S_2} \left[ {b}_{N-1} + {b}_{N-2} {U}_{p}^{N-1} + \cdots + b_1 U_p^2 + b_0 U_p \right].
\end{align}
Here we have used $\prod_{\ell' \in C_1 \cup C_2} {U}_{\ell'} = \prod_{p' \in S_2} {U}_{p'}^{2} \prod_{p' \in S_1 \setminus S_2} {U}_{p'}$ following from (\ref{special}) in the second equality, and (\ref{intp}) in the last equality. In a similar way, we obtain
\begin{equation}
  {Z}_{S_1} \simeq \int \prod_{\ell \in S_1} d {U}_{\ell} \prod_{p \in S_1} \left[ b_0 + {b}_{N-1} {U}_{p}^{N-1} + \cdots + b_2 U_p^2 + b_1 U_p \right].
\end{equation}
In the similar way to the single-winding Wilson loop, the result of integration is given by
\begin{align}
  \langle W(C_1 \cup C_2) \rangle
    &\simeq \frac{ {{b}_{N-2} (\beta)}^{|S_2|} {{b}_{N-1} (\beta)}^{|S_1|-|S_2|} \int \prod_{\ell \in S_1} d {U}_{\ell} }{ {b_0 (\beta)}^{|S_1|} \int \prod_{\ell \in S_1} d {U}_{\ell} } \notag \\
    &= {{c}_{N-2} (\beta)}^{|S_2|} {{c}_{N-1} (\beta)}^{|S_1|-|S_2|} = {c_2 (\beta)}^{|S_2|} {c_1 (\beta)}^{|S_1|-|S_2|} \qquad (N \geqslant 3).
\end{align}
Especially, in the case of $N=3$, due to $c_1 (\beta) = c_2 (\beta)$, we find $\langle W(C_1 \cup C_2) \rangle \simeq {c_1 (\beta)}^{|S_1|}$.

Summarizing the results,
\begin{align}
  \langle W(&C_1 \cup C_2) \rangle =
        \begin{cases}
        {c_1 (\beta)}^{|S_1|-|S_2|} &(N = 2) \\
        {c_1 (\beta)}^{|S_1|} &(N = 3) \\
        {c_2 (\beta)}^{|S_2|} {c_1 (\beta)}^{|S_1|-|S_2|} &(N \geqslant 4)
        \end{cases}.
\end{align}
For $N=2$, $\langle W(C_1 \cup C_2) \rangle$ obeys the difference-of-areas law. 

For $N=3$, $\langle W(C_1 \cup C_2) \rangle$ obeys the max-of-areas law.
This special result in the $N=3$ case is derived from the relation $c_1 (\beta) = c_2 (\beta)$, which holds only in the $N=3$ case and follows from the intrinsic property of the $Z_3$ group in the $Z_N$ theory: ${U}^{2} = {U}^{*}$ for $U \in Z_3$.

For $N \geqslant 4$, the area law depends on $\beta$. In the strong coupling region $\beta \ll 1$, due to $c_1 (\beta) \sim \mathcal{O} (\beta)$ and $c_2 (\beta) \sim \mathcal{O} ({\beta}^{2})$, $\langle W(C_1 \cup C_2) \rangle$ obeys the sum-of-areas law. While in the weak coupling region $\beta \gg 1$, due to $c_1 (\beta) \sim c_2 (\beta)$, $\langle W(C_1 \cup C_2) \rangle$ obeys the max-of-areas law.\\
This result reproduces the areas law falloff obtained by Kato, Shibata, and Kondo in \cite{KSK20} for the double-winding Wilson loop average in the lattice $SU(N)$ pure gauge model, except for the $N$-dependent coefficients reflecting the non-Abelian structure of the group $SU(N)$.
  The max-of-areas law for $N \geqslant 4$ in the weak coupling region is a new result, beyond the region $\beta \ll 1$ where the strong coupling expansion is effective. 
For the spacetime dimension $D>2$, it should be remarked that 
the higher order contributions become the same order as the leading contribution in the weak coupling region due to the asymptotic property $c_n (\beta) \sim 1$. 
Hence, the evaluation of the double-winding Wilson loop average only from the leading contribution is valid only in the strong coupling region $\beta \ll 1$ for $D>2$.

For the original $SU(N)$ theory, in the strong coupling expansion, the candidates of the leading contribution to the coplanar double-winding Wilson loop average are given by the two types of tiling patterns according to the $SU(N)$ group integration formula. 
Although the area $S_1 \setminus S_2$ is tiled by the set of the single plaquettes, in one type of the tiling patterns the area $S_2$ is tiled by the set of  $(N-2)$-fold plaquettes, while in another type the area $S_2$ is tiled by the set of the double plaquettes. See Fig.7 of \cite{KSK20}. 
For $N=2,3$, the former gives the leading contribution (The difference-of-areas law holds for $N=2$ and the max-of-areas law holds for $N=3$), while for $N \geqslant 4$, the latter gives the leading contribution (The sum-of-areas law holds for $N \geqslant 4$). 
This switching giving the leading contribution reproduces the difference between the cases $N=2,3$ and $N \geqslant 4$.
Therefore, the specialness of the $N=3$ case comes from this switching between the two types of contributions.
Note that the right figure in Fig.\ref{fig-bc} for the leading contribution in the case of  $N \geqslant 3$ is $N$-independent in $Z_N$ gauge theory, in contrast to the $SU(N)$ gauge theory. 

Additionally, taking the continuous group limit $N \to \infty$, $b_n (\beta)$ converges to the first-kind modified Bessel function $I_n (\beta)$. From the asymptotic form of the $I_n (\beta)$, $b_n (\beta)$ behaves as
\begin{equation}
  b_n (\beta) \sim \frac{1}{n!} {\left( \frac{\beta}{2} \right)}^{n} \quad (\beta \ll 1), \qquad \frac{{e}^{\beta}}{\sqrt{2 \pi \beta}} \quad (\beta \gg 1),
\end{equation}
and $c_n (\beta)$ have the $N \to \infty$ limit
\begin{equation}
  \lim_{N \to \infty} c_n (\beta) = \frac{I_n (\beta)}{I_0 (\beta)} \sim \frac{1}{n!} {\left( \frac{\beta}{2} \right)}^{n} \quad (\beta \ll 1), \qquad \mathcal{O} (1) \quad (\beta \gg 1).
\end{equation}
Therefore, the areas law for $N \geqslant 4$ persists in the limit $N \to \infty$ i.e. the continuous group $U(1)$. This result suggests the double-winding Wilson loop average in the $U(N)$ lattice gauge model obeys the same area law as that in the $N \geqslant 4$ case of the $SU(N)$ lattice gauge model up to the leading contribution in accord with the center group dominance for the $U(N)$ gauge model, since the center group of $U(N)$ is $U(1)$.\\

\begin{figure}[!h]
\centering
\begin{subfigure}{70mm}
  \centering\includegraphics[width=70mm]{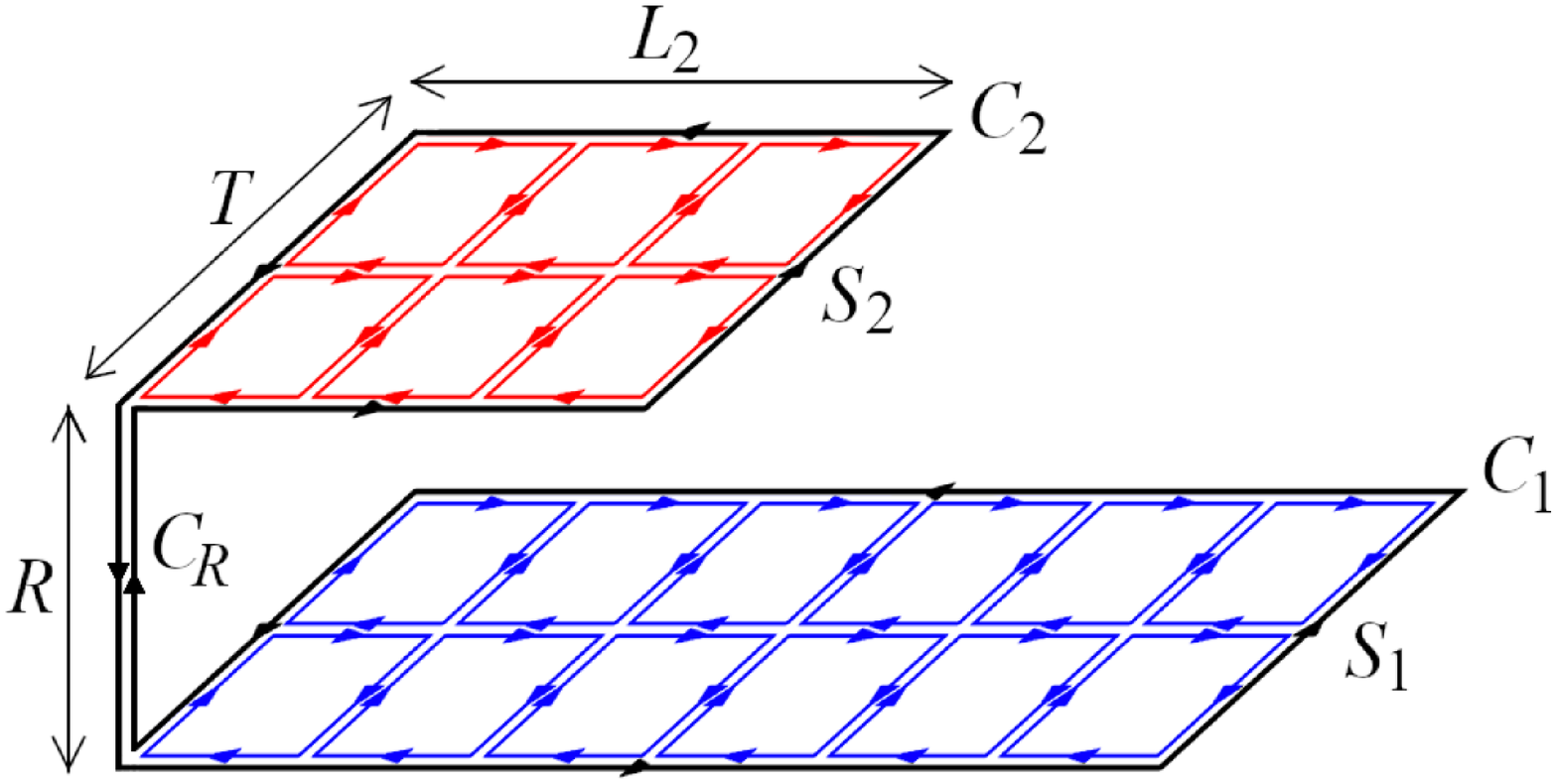}
  \hspace{-10mm} \caption{$diagram (I)$}
  \label{fig-d}
\end{subfigure}
\hspace{10mm}
\begin{subfigure}{70mm}
  \centering\includegraphics[width=70mm]{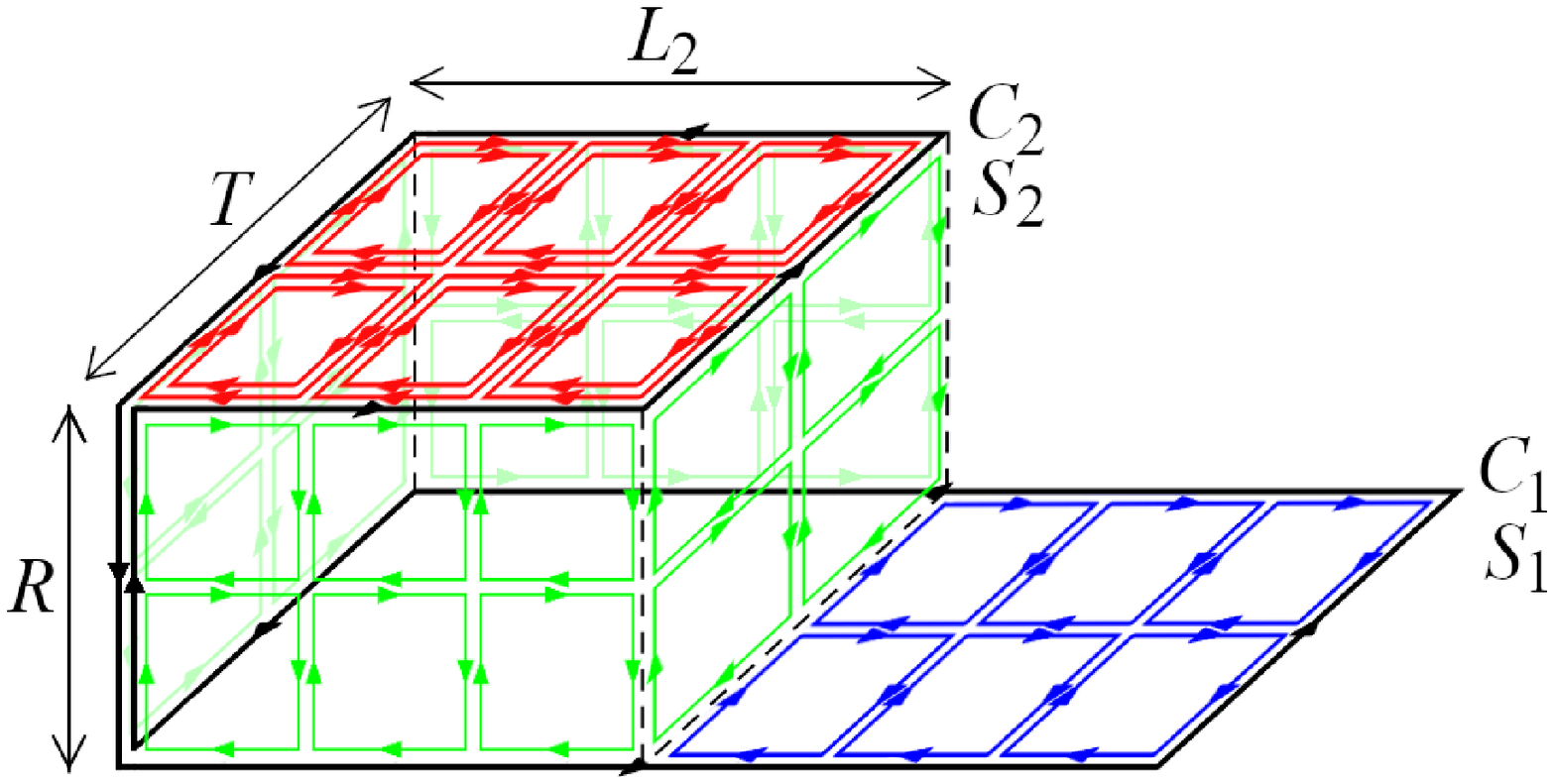}
  \hspace{-10mm} \caption{$diagram (I \hspace{-2pt} I)$}
  \label{fig-e}
\end{subfigure}
\caption{A shifted double-winding Wilson loop, (a) diagram (I), (b)diagram (I\hspace{-1.5pt}I)}
\label{fig-de}
\end{figure}

\noindent
\textit{3.3. A shifted double-winding Wilson loop average in the $Z_N$ pure gauge theory.} \ 
A shifted double-winding Wilson loop is composed of loops $C_1$, $C_2$, and $C_R$. One of the leading contributions is given by the diagram (I) in the left panel of Fig.\ref{fig-de} where the areas $S_1$ and $S_2$ bounded respectively by $C_1$ and $C_2$ are tiled by a minimal set of plaquettes. Note that there is no contribution from $C_R$, because the link variable ${U}_{\ell}$ is Abelian and can be moved to anywhere on the loop $C_1$ and $C_2$ so that it can be arranged to give a trivial result due to ${U}_{\ell} {U}_{\ell}^{*} =1$ before the integration of the link variables on $C_1$ and $C_2$. Therefore, this diagram gives the $R$-independent contribution
\begin{align}
  {\langle W(C_1 \cup C_2) \rangle}_{R \neq 0}^{I}
    &= \langle W(C_1) \rangle \langle W(C_2) \rangle \simeq {c_1 (\beta)}^{|S_1|+|S_2|} \qquad (N \geqslant 2).
\end{align}
Another leading contribution is given by the diagram (I\hspace{-2pt}I) in the right panel of Fig.\ref{fig-de}. 

First, the four sides of a cuboid with the area $2R(L_2+T)$, whose height is $R$ and top is $S_2$, are tiled by plaquettes. After this tiling, the integration is reduced to the case of the coplanar (non-shifted) version. Thus, this diagram gives the $R$-dependent result
\begin{align}
  {\langle W(C_1 \cup C_2) \rangle}_{R \neq 0}^{I \hspace{-2pt} I}
    &\simeq {c_1 (\beta)}^{2R(L_2+T)} \langle W(C_1 \cup C'_2) \rangle \notag \\
    &\simeq\begin{cases}
        {c_1 (\beta)}^{2R(L_2+T)} \cdot {c_1 (\beta)}^{|S_1|-|S_2|} &(N = 2) \\
        {c_1 (\beta)}^{2R(L_2+T)} \cdot {c_1 (\beta)}^{|S_1|} &(N = 3) \\
        {c_1 (\beta)}^{2R(L_2+T)} \cdot {c_2 (\beta)}^{|S_2|} {c_1 (\beta)}^{|S_1|-|S_2|} &(N \geqslant 4)
        \end{cases},
\end{align}
where $C'_2$ denotes a loop $C_2$ after taking the $R \to 0$ limit. Therefore, the total leading contribution for ${\langle W(C_1 \cup C_2) \rangle}_{R \neq 0}$ is composed of the $R$-independent contribution ${\langle W(C_1 \cup C_2) \rangle}_{R \neq 0}^{I}$ and $R$-dependent contribution ${\langle W(C_1 \cup C_2) \rangle}_{R \neq 0}^{I \hspace{-2pt} I}$  :
\begin{align}
  {\langle W(C_1 \cup C_2) \rangle}_{R \neq 0}
    \simeq\begin{cases}
        {c_1 (\beta)}^{|S_1|+|S_2|} + {c_1 (\beta)}^{2R(L_2+T)} \cdot {c_1 (\beta)}^{|S_1|-|S_2|} &(N = 2) \\
        {c_1 (\beta)}^{|S_1|+|S_2|} + {c_1 (\beta)}^{2R(L_2+T)} \cdot {c_1 (\beta)}^{|S_1|} &(N = 3) \\
        {c_1 (\beta)}^{|S_1|+|S_2|} + {c_1 (\beta)}^{2R(L_2+T)} \cdot {c_2 (\beta)}^{|S_2|} {c_1 (\beta)}^{|S_1|-|S_2|} &(N \geqslant 4)
        \end{cases}.
\end{align}
This result reproduces the $R$-dependence of the shifted double-winding Wilson loop average obtained in \cite{KSK20}.

Moreover, we can evaluate the mass gap ${\Delta} (\beta)$ from a shifted double-winding Wilson loop average by considering the case of $S_1 = S_2 = 1$ and $R \gg 1$:
\begin{equation}
  {\langle W(C_1 \cup C_2) \rangle}_{R \neq 0}^{I \hspace{-2pt} I}
    \simeq\begin{cases}
        {c_1 (\beta)}^{4R} = {e}^{4R \ln c_1 (\beta)} &(N = 2) \\
        c_1 (\beta) {c_1 (\beta)}^{4R} = c_1 (\beta) {e}^{4R \ln c_1 (\beta)} &(N = 3) \\
        c_2 (\beta) {c_1 (\beta)}^{4R} = c_2 (\beta) {e}^{4R \ln c_1 (\beta)} &(N \geqslant 4)
    \end{cases},
\end{equation}
according to the relation
\begin{align}
  {\langle W(C_1 \cup C_2) \rangle}_{R \neq 0}^{\textit{conn.}}
  &:= {\langle W(C_1 \cup C_2) \rangle}_{R \neq 0} - \langle W(C_1) \rangle \langle W(C_2) \rangle \notag \\
  &\simeq {\langle W(C_1 \cup C_2) \rangle}_{R \neq 0}^{I \hspace{-2pt} I} \sim C(\beta) {e}^{-R {\Delta} (\beta)} \qquad (R \gg 1),
\end{align}
where $C(\beta)$ is a $\beta$-dependent constant. Therefore, the non-zero mass gap ${\Delta} (\beta)$ is obtained from the result for a shifted double-winding Wilson loop average:
\begin{equation}
  {\Delta} (\beta) = 4 \ln \frac{1}{c_1 (\beta)} > 0 \qquad (0 \leqslant \beta < \infty).
\end{equation}

\noindent
\textit{4. A rigorous result on the center group dominance.} \ 
Finally, we want to mention the works which have focused the role of the center group in quark confinement in the lattice gauge theory from our viewpoint. 
Fr\"{o}hlich has shown \cite{Frohlich79} that
the Wilson loop average at the coupling constant $\beta$ in the lattice non-Abelian gauge theory with the gauge group $G$  is bounded from above by the same Wilson loop average  at the coupling constant $2{\rm dim}(G)\beta$ in the lattice Abelian gauge theory which is obtained by restricting the variables to the center group $Z(G)$ in the same spacetime dimension.  The inequality is explicitly written by using our notations as 
\begin{align}
 |\langle W_{R(G)}(C) \rangle_{G}(\beta)| \le 2 {\rm tr}({\bf 1}) \langle W_{R(Z(G))}(C) \rangle_{Z(G)}(2{\rm dim}(G)\beta),
\label{Frohlich-inequality}
\end{align}
where ${\rm tr}_{R}({\bf 1})$ is the trace of the unit element in the representation $R$ of $G$ and ${\rm dim}(G)$ denotes the dimension of $G$. 
Actually, this inequality was obtained for the gauge-Higgs model where the Higgs scalar field is in the representation that is trivial on $Z(G)$, which is in particular satisfied in the case of the pure gauge theory without the Higgs scalar field. 
Using this inequality, it has been shown that every Abelian and non-Abelian  lattice Higgs theory in the two-dimensional spacetime permanently confines fractionally charged static quarks and that every $U(N)$ $(N=1,2,3,...)$ lattice Yang-Mills theory in the three-dimensional spacetime permanently confines static quarks. 
Similar results are also obtained by Mack and Petkova \cite{MP79} for the $SU(2)$ lattice Yang-Mills theory in the three-dimensional spacetime.

We use the inequality within the pure gauge theory, since we discuss only the pure gauge theory in this paper and the gauge-scalar model is the subject to be tackled in a subsequent paper. 
By examining the proof given in \cite{Frohlich79}, we find that the similar inequality holds also for the double-winding Wilson loop as 
\begin{align}
 |\langle W_{R(G)}(C_1 \cup C_2) \rangle_{G}(\beta)| \le 2 {\rm tr}({\bf 1}) \langle W_{R(Z(G))}(C_1 \cup C_2) \rangle_{Z(G)}(2{\rm dim}(G)\beta).
\label{Frohlich-inequality}
\end{align}
This inequality implies that the area law falloff of the double-winding Wilson loop average in the lattice non-Abelian gauge theory with the gauge group $G$ follows from that in the lattice Abelian gauge theory with the center gauge group $Z(G)$.
In two and three dimensional spacetime, the area law falloff holds for any gauge coupling constant for the Abelian gauge theory, while in the four-dimensional spacetime the area law falloff must hold only in the strong coupling phase because the weak coupling region must be the deconfinement phase (free charge phase for the $Z_N$ group and the Coulomb phase for  the $U(1)$ group). For the non-Abelian gauge theory in the four-dimensional spacetime, therefore, this equality can be used to study the area law falloff only in the strong gauge coupling region. Anyway, the results obtained by explicit calculations in this paper are consistent with this rigorous result.

\noindent
\textit{5. Conclusion and discussion.} \ 
We studied the area law falloff of the double-winding Wilson loops in the lattice $Z_N$ pure gauge model, where the gauge group is the center group of the original gauge group $SU(N)$. 
First, we introduced the lattice $Z_N$ pure gauge model, and applied the character expansion to the weight ${e}^{S_G [U]}$ to investigate the area law beyond the region where the strong coupling expansion works.
Next, we confirmed that the ordinary single-winding Wilson loop average $\langle W(C) \rangle$ obeys the ordinary area law up to the leading contribution, which does not depend on the dimensionality $D$.

Moreover, we evaluated the $N$-dependence of the area law falloff for the coplanar double-winding Wilson loop average ${\langle W(C_1 \cup C_2) \rangle}_{R=0}$ up to the leading contribution. We obtained the difference-of-areas law for $N=2$, the max-of-areas law for $N=3$, and discovered a new $\beta$-dependent result for $N \geqslant 4$ that the sum-of-areas law in the strong coupling region $\beta \ll 1$, and the max-of-areas law in the weak coupling region $\beta \gg 1$. This result reproduces the area law falloff in the lattice $SU(N)$ gauge model obtained in \cite{KSK20}.
We also checked the continuous group limit $N \to \infty$, the area law of the double-winding Wilson loops for $N \geqslant 4$ persists in the lattice $U(1)$ gauge model. This result suggests that the coplanar double-winding Wilson loop average in the lattice $U(N)$ gauge model and the lattice $SU(N) \ (N \geqslant 4)$ gauge model obeys the same area law up to the leading contribution, since $U(1)$ is the center group of $U(N)$.

In addition, we investigated the shifted double-winding Wilson loop average ${\langle W(C_1 \cup C_2) \rangle}_{R \neq 0}$. The total leading contribution is composed of a $R$-independent contribution and a $R$-dependent one. Our result reproduces the $R$-dependent behavior of the shifted double-winding Wilson loop average in the lattice $SU(N)$ gauge model obtained in \cite{KSK20}. We also evaluated the mass gap from this leading result, and obtained the $\beta$-dependent non-zero mass gap $\Delta (\beta)$ for any finite value of $\beta$.

From the above results, we confirmed the center group dominance in reproducing the $N$-dependent area law falloffs of the double-winding Wilson loop averages in the lattice $SU(N)$ gauge theory, as suggested from the rigorous result of Fr\"ohlich \cite{Frohlich79}. 
The result for the double-winding Wilson loop average for the pure gauge model presented in this paper is exact in the whole gauge coupling region only for $D=2$, while it is exact only in the strong gauge coupling region for $D > 2$ which excludes the weak gauge coupling region for $D=4$.

In order to obtain the result valid even in the weak gauge coupling region for $D=4$, we investigate the lattice gauge-scalar model in a subsequent paper. This model allows the analytically connected region \cite{OS78,FS79} between the confinement region ($0 \leqslant \beta \ll 1 , K \ll1$) and the Higgs region ($\beta \gg 1 , K_c \leqslant K < \infty$) which includes the weak gauge coupling region (off the pure gauge region) even if the deconfinement phase exists in the weak gauge coupling region in the pure gauge theory for $D=4$. Therefore, the Abelian gauge dynamics can be relevant in understanding the confinement in the non-Abelian gauge theory for $D=4$.
We will discuss the double-winding Wilson loop averages in the lattice $SU(N)$ gauge scalar theory from this viewpoint in a subsequent paper \cite{IK21b}, by performing the cluster expansion adopted in \cite{OS78,FS79} to estimate the expectation value of the Wilson loop operator which is valid in a specific parameter region of analyticity.

Finally, it should be remarked that the  {center group dominance} derived by Fr\"{o}hlich and the \textit{Abelian dominance} proposed by 't Hooft \cite{tHooft81} is totally different from each other from the physical and mathematical points of view.
The {center group dominance} follows from an inequality between the two expectation values of the Wilson loop operator in the $SU(N)$ and the $Z_N$ gauge theories, which is derived without any gauge fixing procedure in the ordinary framework of lattice gauge theory. Therefore, this result holds irrespective of the choice of gauge fixing condition. Although the area law of the non-Abelian Wilson loop average follows from that of the Abelian Wilson loop average, this is free from the mechanism for the area law in the Abelian gauge theory with the center gauge group. Therefore we can (and must) investigate the gauge-invariant mechanism for confinement without any assumptions afterwards. On the other hand, the {Abelian dominance} of 't Hooft comes after the Abelian projection, which is nothing but the partial gauge fixing for extracting the Abelian gauge theory from the non-Abelian gauge theory.  Moreover, confinement follows from the hypothetical \textit{dual superconductor} vacuum generated by the condensation of magnetic monopoles as the specific topological defects associated with this partial gauge fixing.


\section*{Acknowledgment}

This work was  supported by Grant-in-Aid for Scientific Research, JSPS KAKENHI Grant Number (C) No.19K03840.


%

\vspace{0.2cm}
\noindent


\let\doi\relax


\end{document}